\documentclass[12pt]{article}
\usepackage[left=2.5cm,top=2.50cm,right=2.5cm,bottom=2.50cm]{geometry}
\usepackage{mathrsfs}
\usepackage{amsmath,amssymb,latexsym,color,cancel,graphicx,bbm,colortbl}
\usepackage[english]{babel}
\usepackage[latin1]{inputenc}
\usepackage{ragged2e}
\usepackage{cite}
\begin{document}
\date{}

\title{Algebraic solution and coherent states for the Dirac oscillator interacting with a topological defect}
\author{M. Salazar-Ram\'irez\footnote{{\it E-mail address:} escomphysics@gmail.com}, D. Ojeda-Guill\'en, \\ A. Morales-Gonz\'alez, V.H. Garc\'ia-Ortega} \maketitle

\begin{minipage}{0.9\textwidth}
\small Escuela Superior de C\'omputo, Instituto Polit\'ecnico Nacional,
Av. Juan de Dios B\'atiz esq. Av. Miguel Oth\'on de Mendiz\'abal, Col. Lindavista,
Delegaci\'on Gustavo A. Madero, C.P. 07738, Ciudad de M\'exico, Mexico.\\

\end{minipage}

\begin{abstract}

In this work we study and exactly solve the Dirac oscillator with three different topological defects, namely the cosmic string spacetime($\Lambda_\mp$), the magnetic cosmic string spacetime($\Theta_\mp$) and the cosmic dislocation spacetime ($\Pi_\mp$). Moreover, we show that the radial part of this problem possess an $SU(1,1)$ symmetry. From this, we obtain the wave functions and their respective energy spectrum by means of the Schr\"odinger factorization. Also, we compute the radial coherent states of the eigenfunctions of each problem and their time evolution.\\

\end{abstract}

%PACS: \\
%Keywords:

\section{Introduction}

Nowadays we know that the big bang model suggests that the universe was very hot and dense, and at these high temperatures we would expect the full symmetry to be manifest. As the universe expanded and cooled, it would have undergone a series of phase transitions at which the symmetry was broken. At these phase transitions topologically stable defects may have been formed \cite{Ecope}.

Topological defects are of great importance in many areas of physics such as condensed matter, particles and cosmology. In condensed matter, the topological defects are used to describe systems such as superfluids and superconductors, in models of particle physics to catalyze many unusual effects, and are of great importance in cosmology, since they transport energy that leads to an extra attractive gravitacional force and as seeds of cosmic structures\cite{Davis}.

The Dirac oscillator interacting with topological defects has been widely studied from different points of view. In Ref. \cite{Bakke1}, the influence of the Aharonov-Casher effect on the Dirac oscillator is studied in three different scenarios of general relativity: the Minkowski spacetime, the cosmic string spacetime and the cosmic dislocation spacetime. In Ref. \cite{Bakke2}, the authors study the influence of noninertial effects on the Dirac oscillator in the cosmic string spacetime background. Furthermore, in this work the behavior of the oscillator frequency in a noninertial system that allows to obtain relativistic bound state solutions is discussed.

The Dirac oscillator has also been studied under the influence of nonlinearial effects of a rotating frame in the cosmic string spacetime \cite{Bakke3}. It has been shown that both noninertial effects and the topology of the cosmic string spacetime restrict the physical region of the spacetime where the quantum particle can be placed. The authors also discuss two different cases of bound states of the Dirac equation by analyzing the behavior of the Dirac oscillator frequency. Carvalho et. al. \cite{Carv} solve the Dirac oscillator in the background spacetimes of a cosmic string, of a magnetic cosmic string, and of a cosmic dislocation. In Ref. \cite{Msalazar}, we study a relativistic quantum particle in cosmic string spacetime in the presence of a magnetic field and a Coulomb-type scalar potential.

The factorization method is an alternative way to solve quantum mechanical problems, which some times result to be less complicated than the standard analytical methods. Since Infeld and Hull \cite{Infh1, Infh2} introduced the factorization method in a systematic way by classifying different class of potentials, a large number of quantum systems could be solved \cite{Andri, Spiri, M.salazar2, M.salazar3}. The importance of these factoring methods lies in the fact that if the Schr\"odinger equation is factorizable, the energy spectrum and the eigenfunctions are obtained in an algebraic way. Moreover, the operators constructed from these methods are related to compact and non-compact Lie algebras.

On the other hand, the theory of coherent states arose in order to make comparisons between the classical and quantum theories. These states were introduced by Schr\"odinger \cite{Sch1} in 1926 as the most classical ones of the harmonic oscillator, i.e., those of minimal uncertainty which are not deformed when evolve in time along a classical trajectory. In 1963, Glauber,\cite{Glauber} Klauder,\cite{Klauder1,Klauder2} and Sudarshan\cite{Sudarshan} reintroduced the coherent states for its application in quantum optics. Glauber defined the coherent states as the eigenstates of the annihilation operator of the harmonic oscillator and are related to those the Heisenberg-Weyl group.

The coherent states for the one-dimensional harmonic oscillator were generalized by introducing the displaced number states or number coherent states of the harmonic oscillator. Boiteux and Levelut defined these states by applying the Weyl operator to any excited state $|n\rangle$ and they called them semicoherent states \cite{Boiteux}. Later, Roy and Singh \cite{Roy}, Satyanarayana \cite{Satyanarayana}, and Oliveira et. al. \cite{Kim} gave a detailed study of the properties of these states. Some years later, Nieto \cite{Nieto} derived the most general form of these states.

The Perelomov coherent states were extended by Gerry, who studied the $SU(1,1)$ number coherent states \cite{Gerry0}. Gerry defined these states as the action of the displacement operator onto any $SU(1,1)$ excited state and obtained a general form of these states in terms of the Bargmann V functions. Moreover, he showed that these states are the eigenfunctions of the degenerate parametric amplifier, by an appropriate choice of the coherent state parameters.

In the present paper we study the Dirac oscillator with three different topological defects, namely the cosmic string spacetime, the magnetic cosmic string spacetime and the cosmic dislocation spacetime. In Section 2, we obtain the uncoupled second-order matrix differential equations satisfied by the radial components. In Section 3, we apply the Schr\"odinger factorization method to one of the uncoupled equations and show that the radial part of this problem possess an $SU(1,1)$ symmetry. Then, we use the theory of irreducible representations of the $\mathfrak{su(1,1)}$ Lie algebra to obtain the energy spectrum and the eigenfunctions of our problem. In Section 4, we compute the explicit expression of $SU(1,1)$ Perelomov coherent states for these eigenstates. Finally, we give some concluding remarks.

\section{Dirac oscillator in the the cosmic string spacetime}

The metric tensor for the cosmic string spacetime in cylindrical coordinates is defined by the line element
 \begin{equation}
ds^2=-dt^2+d\rho^2+\alpha^2\rho^2d\phi^2+dz^2,
\end{equation}
where the coordinates $(t,z)\in(-\infty,\infty)$, $\rho<\infty$, the angular variable $\phi\in[0,2\pi]$, $\alpha=1-4\mu$ is the deficit angle associated with the conical geometry of the string and $\mu$ is the linear mass density of the string in natural units.
The Dirac equation for this problem can be written as
\begin{equation}\label{ECUD}
\left[i\gamma^{\mu}(x)-i\gamma^{\mu}(x)\Gamma_\mu(x)-m\right]\Psi(t,x)=0,
\end{equation}
where $\gamma^{\mu}(x)$ are the generalized Dirac matrices, which satisfy the relation $\gamma^{\mu}\gamma^{\nu}+\gamma^{\nu}\gamma^{\mu}=2g^{\mu\nu}$. Notice that in equation (\ref{ECUD}), $\gamma^{\mu}(x)\Gamma_\mu(x)$ is the
correction introduced by the conical geometry of the defect. The $\gamma^{\mu}(x)$ matrices can be written in terms of the standard Dirac matrices $\gamma^{a}$ in Minkowski spacetime as
\begin{equation}
\gamma^{\mu}(x)=e_{(a)}^{\mu}(x)\gamma(a),
\end{equation}
where the tetrad basis $e_{(a)}^{\mu}(x)$ satisfies the relation
\begin{equation}
e_{\mu}^{(a)}(x)e_{\nu}^{(b)}(x)\eta_{ab}=2g_{\mu\nu}.
\end{equation}
The tetrad inverse $e_{(a)}^{\mu}(x)$ in the cosmic string spacetime is written as \cite{Carv}
\begin{equation}\label{matte}
e_{(a)}^{\mu}=\begin{pmatrix}
1 & 0 & 0 & 0 \\
0 & \cos\phi & \sin\phi & 0\\
0 & -\frac{\sin{\phi}}{\alpha\rho} & \frac{\cos{\phi}}{\alpha\rho} & 0\\
0 & 0 & 0 & 1
\end{pmatrix}.
\end{equation}
Therefore, for this representation the Dirac's matrices $\gamma^{\mu}(x)$ obey the relations
\begin{align}\nonumber
\gamma^{0}=&\gamma^{1}, \hspace{0.5cm} \gamma^{3}=\gamma^{z},  \hspace{0.5cm}\gamma^{\rho}=\cos\phi\gamma^{1}+\sin\phi\gamma^{2},\\\label{Gamas}
\gamma^{2}=&\frac{\gamma^{\phi}}{\alpha\rho}, \hspace{0.5cm} \gamma^{\phi}=-\sin\phi\gamma^{1}+\cos\phi\gamma^{2}.
\end{align}
The term $\gamma^{\mu}(x)\Gamma_\mu(x)$ introduced in equation (\ref{ECUD}), can be calculated by using the one-form basis
\begin{align}\nonumber
e^{0}=&dt, \hspace{0.5cm} e^{3}=dz, \hspace{0.5cm} e^{1}=\cos\phi d\rho-\alpha\rho\sin\phi d\phi,\\\label{basis}
e^{2}=&\sin\phi d\rho+\alpha\rho\cos\phi d\phi.
\end{align}
The nonnull components of the spin connection $\omega_{12}$ can be computed with the help of the Maurer-Cartan structure equation $de^a+\omega_b^a\wedge e^b=0$ to obtain $\omega_{12}=-\omega_{21}=(1-\alpha)d\phi$. Thus, the spin connection matrix $\omega_{ab}=\omega_{\mu\hspace{0.1cm}ab}dx^{\mu}$ is
\begin{equation}\label{matte2}
\omega_{\phi\hspace{0.1cm}ab}=\begin{pmatrix}
0 & 0 & 0 & 0 \\
0 & 0 & 1-\alpha & 0\\
0 & -\left(1-\alpha\right) & 0 & 0\\
0 & 0 & 0 & 0
\end{pmatrix}.
\end{equation}
Since $\Gamma_\mu(x)=-\frac{1}{2}\omega_{\mu} ab\frac{\Sigma^{ab}}{2}$ with $\Sigma^{ab}=\frac{1}{2}\left[\gamma^a, \gamma^b\right]$ and by using equations (\ref{Gamas}) and (\ref{matte2}) we obtain that the curvature correction is given by the expression
\begin{equation}\label{Curvacor}
\gamma^{\mu}(x)\Gamma_\mu(x)=\frac{1-\alpha}{2\alpha\rho}\gamma^\rho.
\end{equation}
In order to introduce the Dirac oscillator term $im\omega\beta\rho$, we make the change $\partial_\rho\rightarrow\partial_\rho+m\omega\beta\rho$ in equation (\ref{ECUD}). Thus, by using equations (\ref{Gamas}) and (\ref{Curvacor}), the Dirac oscillator problem in the background of the comic string spacetime can be written as \cite{Carv}
\begin{equation}
\left[-i\gamma^t\partial_t+i\gamma^\rho\left(\partial_\rho-\frac{1-\alpha}{2\alpha\rho}+m\omega\beta\rho\right)+i\frac{\gamma^\phi\partial_\phi}{\alpha\rho}+i\gamma^z\partial_z-m\right]\Psi=0.\label{ECUDir2}
\end{equation}
If we assume temporal independence and rotational symmetry of the defect around $z$ axis, we can propose the Dirac wave function as
\begin{equation}
 \Psi=e^{-iEt+i\left(l+1/2-\Sigma^3/2\right)\phi+ikz}\begin{bmatrix}
\chi(\rho)\\
\phi(\rho)
\end{bmatrix}.
\end{equation}
By using the property $\gamma^\phi\Sigma^3=i\gamma^\rho$, we can write equation (\ref{ECUDir2}) as follows
\begin{equation}\label{ECUDBA}
\left[\beta^2E+i\beta\gamma^\rho\left(\partial_\rho+\frac{1}{2\rho}+m\omega\beta\rho\right)-\beta\gamma^\phi\frac{l+1/2}{\alpha\rho}-\beta\gamma^zk-\beta m\right]\begin{bmatrix}
\chi(\rho)\\
\phi(\rho)
\end{bmatrix}=0.
\end{equation}
On the other hand, equation (\ref{Gamas}) let us obtain the following relationships of the term $\beta\gamma^\mu$
\begin{align}
\beta\gamma^\rho=&\cos\phi\begin{pmatrix}
0 & \sigma^1 \\
\sigma^1 & 0
\end{pmatrix}+\sin\phi\begin{pmatrix}
0 & \sigma^2 \\
\sigma^2 & 0
\end{pmatrix},\\
\beta\gamma^\rho\beta=&\cos\phi \begin{pmatrix}
0 & -\sigma^1 \\
-\sigma^1 & 0
\end{pmatrix}+\sin\phi\begin{pmatrix}
0 & -\sigma^2 \\
-\sigma^2 & 0
\end{pmatrix},\\
\beta\gamma^\phi=&-\sin\phi\begin{pmatrix}
0 & \sigma^1 \\
\sigma^2 & 0
\end{pmatrix}+\cos\phi \begin{pmatrix}
0 & -\sigma^2 \\
-\sigma^1 & 0
\end{pmatrix},\\
\Sigma=&\begin{pmatrix}
0 & -\sigma^2 \\
-\sigma^1 & 0
\end{pmatrix}, \hspace{0.5cm}\beta\gamma^z=\begin{pmatrix}
0 & \sigma^3 \\
\sigma^3 & 0
\end{pmatrix}.
\end{align}

Therefore, with the help of these matrices we can compute the coupled equations of the spinor components $\phi$ and $\chi$
\begin{align}\label{acop1}
&\left(E-m\right)\chi+\left[i\left(\cos\phi\sigma^1+\sin\phi\sigma^2\right)\left(\partial_\rho+\frac{1}{2\rho}-m\omega\beta\rho\right)-\frac{\lambda}{\rho}\left(-\sin\phi\sigma^1+\cos\phi\sigma^2\right)-k\sigma^3\right]\phi=0,\\\label{acopl2} &\left(E+m\right)\phi+\left[i\left(\cos\phi\sigma^1+\sin\phi\sigma^2\right)\left(\partial_\rho+\frac{1}{2\rho}+m\omega\beta\rho\right)-\frac{\lambda}{\rho}\left(-\sin\phi\sigma^1+\cos\phi\sigma^2\right)-k\sigma^3\right]\chi=0. \end{align}

The uncoupled equation for $\chi$ can be obtained by multiplying the equation (\ref{acop1}) by the term $\left(E+m\right)$. Thus, from equation (\ref{acopl2}) we obtain
\begin{align}\label{acop2}\nonumber
&\left(E^2-m^2\right)\chi-\left[i\left(\cos\phi\sigma^1+\sin\phi\sigma^2\right)\left(\partial_\rho+\frac{1}{2\rho}-m\omega\beta\rho\right)-\frac{\lambda}{\rho}\left(-\sin\phi\sigma^1+\cos\phi\sigma^2\right)-k\sigma^3\right]\\  &\times\left[i\left(\cos\phi\sigma^1+\sin\phi\sigma^2\right)\left(\partial_\rho+\frac{1}{2\rho}+m\omega\beta\rho\right)-\frac{\lambda}{\rho}\left(-\sin\phi\sigma^1+\cos\phi\sigma^2\right)-k\sigma^3\right]\chi=0.
\end{align}
If we apply an analogous procedure to the other spinor component $\phi$ we obtain that its uncoupled differential equation is
\begin{align}\label{acop3}\nonumber
&\left(E^2-m^2\right)\phi-\left[i\left(\cos\phi\sigma^1+\sin\phi\sigma^2\right)\left(\partial_\rho+\frac{1}{2\rho}-m\omega\beta\rho\right)-\frac{\lambda}{\rho}\left(-\sin\phi\sigma^1+\cos\phi\sigma^2\right)-k\sigma^3\right]\\  &\times\left[i\left(\cos\phi\sigma^1+\sin\phi\sigma^2\right)\left(\partial_\rho+\frac{1}{2\rho}-m\omega\beta\rho\right)-\frac{\lambda}{\rho}\left(-\sin\phi\sigma^1+\cos\phi\sigma^2\right)-k\sigma^3\right]\phi=0.
\end{align}
The matrix differential equation for the spinor component $\chi$ can be written as \cite{Carv}
\begin{align}
&\frac{d^2\chi(\rho)}{d\rho^2}+\frac{1}{\rho}\frac{d\chi(\rho)}{d\rho}-\frac{1}{\rho^2}\left(\frac{1}{4}+i\lambda\sigma^{1}\sigma^{2}+\lambda^2\right)\chi(\rho)
-m\omega^2\rho^2\chi(\rho)+\left(E^2-m^2+m\omega-k^2\right)\chi(\rho)\nonumber\\ &-2m\omega\left[i\lambda\sigma^{1}\sigma^{2}+ik(\rho\cos{\phi}\sigma^{1}\sigma{3}
+\rho\sin{\phi}\sigma^{2}\sigma^{3})\right]\chi(\rho)=0.\label{desac}
\end{align}
An analogous expression can be obtained for the spinor component $\chi$.

Therefore, we have computed the matrix differential equations of the Dirac oscillator radial spinor components $\phi$ and $\chi$ in the comic string spacetime.

\section{Algebraic solution using the Schr\"odinger factorization method}

In this Section we shall obtain the exact solution of the Dirac oscillator in the cosmic string spacetime by means of the Schr\"odinger factorization method. It can be shown that the last term of equation (\ref{desac}) can be written as \cite{Carv}
\begin{equation}
\left[i\lambda\sigma^1\sigma^2+ik\left(\rho\cos\phi\sigma^1\sigma^3+\rho\sin\phi\sigma^2\sigma^3\right)\right]=-2\overrightarrow{S}\cdot\overrightarrow{L},
\end{equation}
where it has been used the definition of the scalar product $\overrightarrow{S}\cdot \overrightarrow{L}$ in cylindrical coordinates and the relationship $\overrightarrow{S}=\frac{\overrightarrow{\sigma}}{2}$. Now, considering a tridimensional planar system were the term $\overrightarrow{S}\cdot\overrightarrow{L}$ has an eigenvalue $\left(\frac{l+1/2}{2\alpha}\right)$, the equation (\ref{desac}) can finally be written in a compact form as \cite{Carv}
\begin{align}\label{second} \left[\frac{d^2}{d\rho^2}+\frac{1}{\rho}\frac{d}{d\rho}-\frac{\Lambda_\mp^2}{\rho^2}-m^2\omega^2\rho^2+\gamma_\pm\right]\Psi=0,
\end{align}
with $\Psi=\begin{pmatrix}
\chi\\
\phi
\end{pmatrix}$, and where
\begin{align}\label{cmvar}
\Lambda_\mp=&\frac{l+\frac{1}{2}}{\alpha}\mp\frac{1}{2},\\
\gamma_\pm=&E^2-m^2+2m\omega\left(\Lambda_\pm\right)-k^2.
\end{align}

In addition to the topological defect of the cosmic string spacetime, we can consider two more defects, namely the magnetic cosmic string background and the cosmic dislocation spacetime. The magnetic cosmic string ($\Theta_\mp$) is characterized by a magnetic potential vector given by
\begin{equation}
\overrightarrow{A_\mu}=i\frac{\Phi_B}{2\pi\alpha\rho}\widehat{e}_\phi,
\end{equation}
which gives a flux tube coinciding with the cosmic string and the z axis. 

On the other hand, the Dirac oscillator in cosmic dislocation spacetime ($\Pi_\mp$) is described in cylindrical coordinates by the following metric \cite{Galstov}
\begin{equation}
ds^2=-dt^2+d\rho^2+\alpha^2\rho^2d\phi^2+\left(dz+J^zd\phi\right)^2,
\end{equation}
with $\rho\geq0$ and $0\leq\phi\leq2\pi$. The parameter $J^z$ is related to the torsion source and if we set $J^z=0$, the metric  of the cosmic string spacetime is recovered. 

It should be noted that the differential equations of all topological defects considered in this work, the cosmic string spacetime ($\Lambda_\mp$), the magnetic cosmic string ($\Theta_\mp$) and the cosmic dislocation spacetime ($\Pi_\mp$) can be written in a single expression. Therefore, equation (\ref{second}) can be written in a compact form as

\begin{align}\label{difpri} \left[\frac{d^2}{d\rho^2}+\frac{1}{\rho}\frac{d}{d\rho}-\frac{\Lambda_\mp^2,\Theta_\mp^2,\Pi_\mp^2}{\rho^2}-m^2\omega^2\rho^2+\Gamma_\pm\right]\Psi=0,
\end{align}

with
\begin{align}\label{cmvar2}
\Theta_\mp:=&\frac{l+\frac{1}{2}+e\Phi_B/2\pi}{\alpha}\mp\frac{1}{2},\\
\Pi_\mp:=&\frac{l+\frac{1}{2}+e\Phi_B/2\pi-kJ^z}{\alpha}\mp\frac{1}{2},\\\label{Gamm}
\Gamma_\pm=&E^2-m^2+2m\omega\left(\Lambda_\pm,\Theta_\pm,\Pi_\pm\right)-k^2.
\end{align}

Then, in order to solve the general equation (\ref{difpri}), we proceed to remove the first derivative by making the change of variable $\begin{pmatrix}
\chi\\
\phi
\end{pmatrix}=\frac{1}{\sqrt{\rho}}\begin{pmatrix}
F\\
G
\end{pmatrix}$ to obtain

\begin{equation}\label{secondFac}
\left(-\rho^2\frac{d^2}{d\rho^2}+m^2\omega^2\rho^4-\Gamma_\pm\rho^2\right)\begin{pmatrix}
F\\
G
\end{pmatrix}=\left(\frac{1}{4}-\left[\Lambda_\mp^2,\Theta_\mp^2,\Pi_\mp^2\right]\right)\begin{pmatrix}
F\\
G
\end{pmatrix}.
\end{equation}
We will focus on the second order differential equation for the radial function $F$ of equation (\ref{secondFac}), since under the change $\left(\Lambda_\mp,\Theta_\mp,\Pi_\mp\right)\rightarrow \left(\Lambda_\pm,\Theta_\pm,\Pi_\pm\right)\mp1$  we obtain the radial function for $G$.

From equation (\ref{secondFac}) we construct the $\mathfrak{su(1,1})$ algebra generators by applying the Schr\"odinger factorization. Thus, we propose
\begin{equation}\label{sch}
\left(\rho\frac{d}{d\rho}+\mu\rho^2+\delta\right)\left(-\rho\frac{d}{d\rho}+\varepsilon\rho^2+\lambda\right)F=\sigma{F}.
\end{equation}
Expanding this expression and comparing it with equation (\ref{secondFac}) we obtain that the constants $\mu,\delta, \varepsilon,\lambda$ are
\begin{align}
\mu=&\varepsilon=\pm{m\omega}, \hspace{1.0cm} \lambda=\delta+1=-\frac{\Gamma_+}{2m\omega}-\frac{1}{2},\\
\sigma=&\left(\frac{\Gamma_+}{2m\omega}+1\right)^2-\left[\Lambda_-^2,\Theta_-^2,\Pi_-^2\right].
\end{align}
Therefore, the differential equation for $F$ can be factorized as
\begin{equation}
\left[\mathfrak{B}_\mp\mp1\right]\mathfrak{B}_\pm=\frac{1}{4}\left[\left(\frac{\Gamma_+}{2m\omega}\pm1\right)^2-\left(\Lambda_-^2,\Theta_-^2,\Pi_-^2\right)\right],
\end{equation}
with
\begin{equation}\nonumber
\mathfrak{B}_\pm=\frac{1}{2}\left[\mp\rho\frac{d}{d\rho}+m\omega\rho^2-\frac{1}{2m\omega}\left(E^2-m^2-k^2\right)+
\left[\Lambda_+,\Theta_+,\Pi_+\right]\mp\frac{1}{2}\right],
\end{equation}
the Schr\"odinger operators. Thus, we can introduce the new pair of operators given by
\begin{equation}
\mathfrak{D}_\pm=\frac{1}{2}\left[\mp\rho\frac{d}{d\rho}+m\omega\rho^2\mp\frac{1}{2}\right]-\mathfrak{D}_3,
\end{equation}
where the operator $\mathfrak{D}_3$ is defined from equation (\ref{secondFac}) as
\begin{align}\nonumber\label{oper3}
\mathfrak{D}_3F=&\frac{1}{4m\omega}\left[-\frac{d^2}{d\rho^2}+\frac{\left(\Lambda_-^2,\Theta_-^2,\Pi_-^2\right)-\frac{1}{4}}{\rho^2}+m^2\omega^2\rho^2\right]F\\
=&\frac{1}{4m\omega}\left[\left(E^2-m^2-k^2\right)+\frac{1}{2}\left[\Lambda_+,\Theta_+,\Pi_+\right]\right]F.
\end{align}
It can be shown that these operators close the $\mathfrak{su(1, 1)}$ Lie algebra
\begin{equation}\label{com}
\left[\mathfrak{D}_3, \mathfrak{D}_\pm\right]=\pm\mathfrak{D}_\pm, \hspace{0.5cm} \left[\mathfrak{D}_-,\mathfrak{D}_+\right]=2\mathfrak{D}_3.
\end{equation}
It is well known that the Casimir operator for this algebra, is given by $\mathfrak{C}^2=-\mathfrak{D}_+\mathfrak{D}_-+\mathfrak{D}_3\left(\mathfrak{D}_3-1\right)$, and satisfies the eigenvalue equation
\begin{equation}\label{Cas}
\mathfrak{C}^2F=\left[\left(\Lambda_-^2,\Theta_-^2,\Pi_-^2\right)-1\right]F=k(k-1)F.
\end{equation}
This equality holds due to the representation theory of the $\mathfrak{su(1,1)}$ Lie algebra (see Appendix). Therefore, the relationship between $\left(\Lambda_-,\Theta_-,\Pi_-\right)$ and the quantum number $k$ is obtained by using equation (\ref{Cas}). Moreover, the other group number $n$, can be identified with the radial quantum number $n_r$. By comparison of equations (\ref{oper3}) and (\ref{k0n}) we obtain
\begin{align}\label{numk}
k=&\frac{1}{2}|\Lambda_-,\Theta_-,\Pi_-|+\frac{1}{2}, \hspace{0.5cm} n=n_r,\\
n+k=&n_r+\frac{1}{4}|\Lambda_-,\Theta_-,\Pi_-|+\frac{1}{2},
\end{align}
with $n_r=0,1,2,......$

The energy spectrum for this spinor can be computed from equations (\ref{k0n}), (\ref{oper3}) and (\ref{numk}) to obtain
\begin{equation}
E=m^2+4m\omega\left[n_r+\frac{1}{2}|\Lambda_-,\Theta_-,\Pi_-|- \right.
\frac{1}{2}\left(\Lambda_+,\Theta_+,\Pi_+\right)+\frac{1}{2}\bigg]+k^2.
\end{equation}

We apply a similar procedure for the lower spinor component G. Thus, the pair of operators for this case are
\begin{equation}
\mathfrak{J}_\pm=\frac{1}{2}\left[\mp\rho\frac{d}{d\rho}+m\omega\rho^2\mp\frac{1}{2}\right]-\mathfrak{J}_3,
\end{equation}
where the operator $\mathfrak{J}_3$ is
\begin{align}\label{J3}\nonumber
\mathfrak{J}_3G=&\frac{1}{4m\omega}\left[-\frac{d^2}{d\rho^2}+\frac{\left(\Lambda_+^2,\Theta_+^2,\Pi_+^2\right)-\frac{1}{4}}{\rho^2}+m^2\omega^2\rho^2\right]G\\
=&\frac{1}{4m\omega}\left[\left(E^2-m^2-k^2\right)+\frac{1}{2}\left[\Lambda_-,\Theta_-,\Pi_-\right]\right]G,
\end{align}
and
\begin{equation}
k=\frac{1}{2}|\Lambda_+,\Theta_+,\Pi_+|+\frac{1}{2}.
\end{equation}
In the same way as for the upper spinor, the operators $\mathfrak{J}_\pm, \mathfrak{J}_3$ close $\mathfrak{su(1,1)}$ Lie algebra. The energy spectrum for this spinor is given by
\begin{equation}
E=m^2+4m\omega\left[n_r+\frac{1}{2}|\Lambda_+,\Theta_+,\Pi_+| \right.
-\frac{1}{2}\left(\Lambda_-,\Theta_-,\Pi_-\right)+\frac{1}{2}\bigg]+k^2.
\end{equation}
Dividing equation (\ref{second}) between $-m\omega\rho^2$ and making the change $-m\omega\rho^2 \rightarrow r^2$ we obtain
\begin{equation}\label{second2}
\left(\frac{d}{dr^2}+\frac{\frac{1}{4}-\left(\Lambda_\mp^2,\Theta_\mp^2,\Pi_\mp^2\right)}{r^2}-r^2+\frac{\Gamma_\pm}{m\omega}\right)\begin{pmatrix}
F\\
G
\end{pmatrix}=0.
\end{equation}
Thus, the radial functions $F(r)$ and $G(r)$ can be obtained from the general differential equation\cite{LEB}
\begin{equation}
z''+\left[4n+2\alpha+2-x^2+\frac{\frac{1}{4}-\alpha^2}{x^2}\right]z=0,\label{difgen}
\end{equation}
which has the particular solution
\begin{equation}
z=N_ne^{-x^2/2}x^{\alpha+1/2}L_n^{\alpha}(x^2).
\end{equation}
Therefore, the radial wave functions $F(r)$ and $G(r)$ explicitly are
\begin{align}\label{second3}
\begin{bmatrix}
F(\rho)\\
G(\rho)
\end{bmatrix}=&\frac{2\Gamma\left(n_r+1\right)}{\Gamma\left(n_r+|\Lambda_\mp,\Theta_\mp,\Pi_\mp|+1\right)}e^{\frac{-m\omega\rho^2}{2}}
\left(m\omega\right)^{\frac{|\Lambda_\mp,\Theta_\mp,\Pi_\mp|+\frac{1}{2}}{2}}\rho^{|\Lambda_\mp,\Theta_\mp,\Pi_\mp|+\frac{1}{2}}\\\nonumber
\times &L_{n_r}^{|\Lambda_\mp,\Theta_\mp,\Pi_\mp|}(mw\rho^2),
\end{align}
where the normalization coefficient $N_n$ was computed from the orthogonality of the Laguerre polynomials
\begin{equation}
\int_0^{\infty}e^{-x}x^{\alpha}\left[L_{n}^{\alpha}(x)\right]^2dx=\frac{\Gamma(n+\alpha+1)}{n!}.
\end{equation}
The Sturmian basis for the irreducible unitary representation of the $\mathfrak{su(1,1)}$
Lie algebra in terms of the group numbers $n$, $k_{(\chi,\phi)}=\frac{1}{2}|\Lambda_\mp,\Theta_\mp,\Pi_\mp|+\frac{1}{2}$ for the Dirac oscillator interacting with a topological defect are
\begin{align}
\begin{bmatrix}
\chi(\rho)\\
\phi(\rho)
\end{bmatrix}_{n_r,\Lambda_\mp,\Theta_\mp,\Pi_\mp}=&\left[\frac{2\Gamma\left(n_r+1\right)}{\Gamma\left(n_r+|\Lambda_\mp,\Theta_\mp,\Pi_\mp|+1\right)}\right]^{\frac{1}{2}}
e^{\frac{-m\omega\rho^2}{2}}\left(m\omega\right)^{\frac{|\Lambda_\mp,\Theta_\mp,\Pi_\mp|+\frac{1}{2}}{2}}\\\nonumber
\times &\rho^{|\Lambda_\mp,\Theta_\mp,\Pi_\mp|}L_{n_r}^{|\Lambda_\mp,\Theta_\mp,\Pi_\mp|}(mw\rho^2),
\end{align}
or
\begin{multline}\label{sturm1}
\begin{bmatrix}
\chi(\rho)\\
\phi(\rho)
\end{bmatrix}_{n,k_{(\chi,\phi)}}=\left[\frac{2\Gamma\left(n+1\right)}{\Gamma\left(n+2k_{(\chi,\phi)}\right)}\right]^{\frac{1}{2}}e^{\frac{-m\omega\rho^2}{2}}
\left(m\omega\right)^{k_{(\chi,\phi)}+\frac{1}{2}}\rho^{2k-1} L_{n}^{2k_{(\chi,\phi)}-1}(mw\rho^2).
\end{multline}

\section{$SU(1,1)$ radial coherent states and their time evolution}

From the definition of $SU(1,1)$ Perelomov coherent states \cite{PERL}, we shall construct the relativistic coherent states for the radial functions $\chi$ and $\phi$  by using the Sturmian basis
\begin{equation}
|\zeta\rangle=D(\xi)|k,0\rangle=(1-|\xi|^2)^k\sum_{n=0}^\infty\sqrt{\frac{\Gamma(n+2k)}{n!\Gamma(2k)}}\xi^n|k,n\rangle.
\end{equation}
Here, $D(\xi)$ is the displacement operator and $|k, 0\rangle$ the lowest normalized state. Thus, by applying the operator $D(\xi)$ to the ground states of the functions $\chi$ and $\phi$ , we obtain
\begin{align}\label{sturm2}
\begin{bmatrix}\nonumber
\chi(\rho,\xi)\\
\phi(\rho,\phi)
\end{bmatrix}=&\left[\frac{2\left(1-|\xi|^2\right)^{2k_{(\chi,\phi)}}}{\Gamma\left(2k_{(\chi,\phi)}\right)}\right]^{\frac{1}{2}}e^{\frac{-m\omega\rho^2}{2}}\left(m\omega\right)^{k_{(\chi,\phi)}+\frac{1}{2}}\\
\times &\rho^{2k_{(\chi,\phi)}-1}\sum_{n=0}^\infty\xi^nL_{n}^{2k_{(\chi,\phi)}-1}\left(m\omega\rho^2\right).
\end{align}
The latter sum can be calculated from the Laguerre polynomials generating function
\begin{equation}
\sum_{n=0}^\infty L_n^\nu(x)y^n=\frac{e^{-xy/(1-y)}}{(1-y)^{\nu+1}},
\end{equation}
to obtain that the radial coherent states $\chi(\rho,\xi)$ and $\phi(\rho,\xi)$ can be written as
\begin{align}\label{estcoh}
\begin{bmatrix}
\chi(\rho,\xi)\\
\phi(\rho,\xi)
\end{bmatrix}=\left[\frac{2\left(1-|\xi|^2\right)^{2k_{(\chi,\phi)}}}{\Gamma\left(2k\right)\left(1-\xi\right)^{4k_{(\chi,\phi)}}}\right]^{\frac{1}{2}}\left(m\omega\right)^{k_{(\chi,\phi)}+\frac{1}{2}}
\rho^{2k_{(\chi,\phi)}-1}e^{\frac{m\omega\rho^2}{2}\left(\frac{\xi+1}{\xi-1}\right)}.
\end{align}
The $SU(1,1)$ radial coherent states for the Dirac oscillator interacting with a topological defect in terms of the physical quantum numbers $\Lambda_\mp,\Theta_\mp,\Pi_\mp$ are given by
\begin{equation}\label{estcoh2}
\begin{bmatrix}
\chi(\rho,\xi)\\
\phi(\rho,\xi)
\end{bmatrix}=\left[\frac{2\left(1-|\xi|^2\right)^{|\Lambda_\mp,\Theta_\mp,\Pi_\mp|+1}}{\Gamma\left(|\Lambda_\mp,\Theta_\mp,\Pi_\mp|+1\right)} \frac{\left(m\omega\right)^{|\Lambda_\mp,\Theta_\mp,\Pi_\mp|+\frac{1}{2}}}{\left(1-\xi\right)^{2|\Lambda_\mp,\Theta_\mp,\Pi_\mp|+2}}\right]^{\frac{1}{2}}\rho^{|\Lambda_\mp,\Theta_\mp,\Pi_\mp|}e^{\frac{m\omega\rho^2}{2}\left(\frac{\xi+1}{\xi-1}\right)}.
\end{equation}
In order to obtain the time evolution of these coherent states,  we write equation (\ref{second}) as
\begin{equation}\label{Hgam}
H_r\begin{pmatrix}
F\\
G
\end{pmatrix}=\Gamma_\pm\begin{pmatrix}
F\\
G
\end{pmatrix},
\end{equation}
where
\begin{equation}\label{second4}
H_r=\left(-\rho^2\frac{d^2}{d\rho^2}+\frac{\left[\Lambda_\mp^2,\Theta_\mp^2,\Pi_\mp^2\right]-\frac{1}{4}}{\rho^2}+m^2\omega^2\rho^2\right).
\end{equation}
Thus, from equation (\ref{oper3}), (\ref{J3}) and (\ref{Hgam}) the following relationship can be obtained
\begin{equation}
H_r\begin{pmatrix}
F\\
G
\end{pmatrix}=4m\omega\begin{pmatrix}\mathfrak{D_3}\\
 \mathfrak{J_3}
\end{pmatrix}
\begin{pmatrix}
F\\
G
\end{pmatrix}=\Gamma_\pm\begin{pmatrix}
F\\
G
\end{pmatrix}.
\end{equation}

Using this equation and equations (\ref{oper3}), (\ref{J3}), (\ref{Gamm}) and (\ref{k0n}), we can obtain again the energy spectrum
\begin{equation}
E=m^2+4m\omega\left[n_r+\frac{1}{2}|\Lambda_\mp,\Theta_\mp,\Pi_\mp|
-\frac{1}{2}\left(\Lambda_\pm,\Theta_\pm,\Pi_\pm\right)+\frac{1}{2}\right]+k^2
\end{equation}
The time evolution operator for an arbitrary Hamiltonian is defined as \cite{COH}
\begin{equation}\label{UTEM}
\mathscr{U}(\tau)=e^{-iH_{r}\tau/\hbar}=e^{-4im\omega\mathfrak{D_3}\tau/\hbar},
\end{equation}
where $\tau$ is considered as a fictitious time \cite{GUR,GER}. As it is well known \cite{NOS1},  the time evolution
of the Perelomov coherent states is given by
\begin{equation}\label{PERET}
|\zeta(\tau)\rangle =\mathscr{U}(\tau)|\zeta\rangle=\mathscr{U}(\tau)D(\xi)\mathscr{U}^\dag(\tau)\mathscr{U}(\tau)|k,0\rangle,
\end{equation}
where $D(\xi)=\exp(\xi \mathfrak{D}_{+}-\xi^{*}\mathfrak{D}_{-})$ is the displacement operator and $\xi$ is a complex number. From equation (\ref{k0n}) of Appendix, the time evolution of the state $|k,0\rangle$ is
\begin{equation}\label{evest1}
\mathscr{U}(\tau)|k,0\rangle=e^{-4im\omega k\tau/\hbar}|k,0\rangle.
\end{equation}
The similarity transformation $\mathscr{U}(\tau)D(\xi)\mathscr{U}^\dag(\tau)$ contained in equation (\ref{PERET}) can be computed from the time evolution of the raising and lowering operators $\mathscr{U}^\dag(\tau)\mathfrak{D}_{\pm}\mathscr{U}(\tau)$ . Thus, using the BCH (Baker-Campbell-Hausdorff) formula and equations (\ref{com}) and (\ref{UTEM}) is obtained
\begin{align}
\mathfrak{D}_+(\tau)=&\mathscr{U}^\dag(\tau)\mathfrak{D}_+\mathscr{U}(\tau)=\mathfrak{D}_+e^{4im\omega \tau/\hbar},\\
\mathfrak{D}_-(\tau)=&\mathscr{U}^\dag(\tau)\mathfrak{D}_-\mathscr{U}(\tau)=\mathfrak{D}_-e^{-4im\omega \tau/\hbar}.
\end{align}
Therefore, the similarity transformation $\mathscr{U}(\tau)D(\xi)\mathscr{U}^\dag(\tau)$ can be written as
\begin{align}\label{opd1}
\mathscr{U}(\tau)D(\xi)\mathscr{U}^\dag(\tau)=e^{\xi \mathfrak{D}_+(-\tau)-\xi^*\mathfrak{D}_-(-\tau)}=e^{\xi(-\tau)\mathfrak{D}_+ - \xi(-\tau)^*\mathfrak{D}_-},
\end{align}
with $\xi(t)=\xi e^{\frac{-4im\omega\tau}{\hbar}}$. Now, if we define $\zeta(t)=\zeta e^{\frac{-4im\omega\tau}{\hbar}}$, the time evolution of the displacement operator in its normal form can be expressed as
\begin{equation}\label{opedes}
D(\xi(t))=e^{\zeta(t) \mathfrak{D}_+}e^{\eta \mathfrak{D}_0}e^{-\zeta(t)^*\mathfrak{D}_-}.
\end{equation}
From equations (\ref{evest1}) and (\ref{opedes}), we obtain that the time dependent Perelomov coherent state is
\begin{equation}
|\zeta(t)\rangle=e^{-4im\omega k\tau/\hbar}e^{\zeta(-\tau)\mathfrak{D}_+}e^{\eta \mathfrak{D}_0}e^{-\zeta(-\tau)^*\mathfrak{D}_-}|k,0\rangle.
\end{equation}

This result allows us to compute the time evolution of the coherent state for the Dirac oscillator interacting with a topological defect  in the configuration space
\begin{align}\label{estcoh3}
\begin{bmatrix}
\chi(\rho,\xi)\\
\phi(\rho,\xi)
\end{bmatrix}=&\left[\frac{2\left(1-|\xi|^2\right)^{|\Lambda_\mp,\Theta_\mp,\Pi_\mp|+1}}{\Gamma\left(|\Lambda_\mp,\Theta_\mp,\Pi_\mp|+1\right)} \frac{\left(m\omega\right)^{|\Lambda_\mp,\Theta_\mp,\Pi_\mp|+\frac{1}{2}}}{\left(1-\xi{e}^{4im\omega\tau/\hbar}\right)^{2|\Lambda_\mp,\Theta_\mp,\Pi_\mp|+2}}\right]^{\frac{1}{2}}\rho^{|\Lambda_\mp,\Theta_\mp,\Pi_\mp|}\times\\\nonumber
\times & e^{-4im\omega \left(\frac{1}{2}|\Lambda_\mp,\Theta_\mp,\Pi_\mp|+\frac{1}{2}\right)}e^{\frac{m\omega\rho^2}{2}\left(\frac{\xi{e}^{4im\omega\tau/\hbar}+1}{\xi{e}^{4im\omega\tau/\hbar}-1}\right)}.
\end{align}

\section{Concluding remarks}
In this paper we studied the Dirac oscillator interacting with three topological defects, the cosmic string spacetime, the magnetic cosmic string and the cosmic dislocation spacetime from an algebraic approach. In order to solve our problem, we focused on one of the two uncoupled second order radial equations of the spinor. We showed that the radial part for this problem possess an $SU(1,1)$ symmetry and by using Schr\"odinger factorization we were able to construct the $\mathfrak{su(1,1)}$ algebra generators of each spinor component. We obtained the energy spectrum and eigenfunctions by using the theory of unitary representations for this algebra. From the Sturmian basis of the $\mathfrak{su(1,1)}$ Lie algebra we constructed the Perelomov coherent states for each radial component. Then, we computed the radial relativistic coherent state and its time evolution for the Dirac oscillator interacting with each topological defect.

\section*{Acknowledgments}

This work was partially supported by SNI-M\'exico, COFAA-IPN, EDI-IPN, EDD-IPN, SIP-IPN project number $20180245$.

\renewcommand{\theequation}{A.\arabic{equation}}
\setcounter{equation}{0}
\section*{Appendix. The ${SU(1,1)}$ Group and its coherent states}

Three operators $K_{\pm}, K_0$ close the $\mathfrak{su(1,1)}$ Lie algebra if they satisfy the commutation relations \cite{Vourdas}
\begin{eqnarray}
[K_{0},K_{\pm}]=\pm K_{\pm},\quad\quad [K_{-},K_{+}]=2K_{0}.\label{comm}
\end{eqnarray}
The action of these operators on the basis states (Sturmian states)
$\{|k,n\rangle, n=0,1,2,...\}$ is
\begin{equation}
K_{+}|k,n\rangle=\sqrt{(n+1)(2k+n)}|k,n+1\rangle,\label{k+n}
\end{equation}
\begin{equation}
K_{-}|k,n\rangle=\sqrt{n(2k+n-1)}|k,n-1\rangle,\label{k-n}
\end{equation}
\begin{equation}
K_{0}|k,n\rangle=(k+n)|k,n\rangle,\label{k0n}
\end{equation}
where $|k,0\rangle$ is the lowest normalized state.

The theory of unitary irreducible representations of the $\mathfrak{su(1,1)}$ Lie algebra has been
studied in several works \cite{ADAMS} and it is based on equations (\ref{k+n})-(\ref{k0n}) and (\ref{Cas}).
Thus, a representation of $\mathfrak{su(1,1)}$ algebra is determined by the number $k$.

The $SU(1,1)$ Perelomov coherent states $|\zeta\rangle$ are
defined as \cite{PERL}
\begin{equation}
|\zeta\rangle=D(\xi)|k,0\rangle,\label{defPCS}
\end{equation}
where $D(\xi)=\exp(\xi K_{+}-\xi^{*}K_{-})$ is the displacement
operator and $\xi$ is a complex number. From the properties
$K^{\dag}_{+}=K_{-}$ and $K^{\dag}_{-}=K_{+}$ it can be shown that
the displacement operator possesses the property
\begin{equation}
D^{\dag}(\xi)=\exp(\xi^{*} K_{-}-\xi K_{+})=D(-\xi),
\end{equation}
and the so called normal form of the displacement operator is given by
\begin{equation}
D(\xi)=\exp(\zeta K_{+})\exp(\eta K_{0})\exp(-\zeta^*
K_{-})\label{normal},
\end{equation}
where $\xi=-\frac{1}{2}\tau e^{-i\varphi}$, $\zeta=-\tanh
(\frac{1}{2}\tau)e^{-i\varphi}$ and $\eta=-2\ln \cosh
|\xi|=\ln(1-|\zeta|^2)$ \cite{GER}. By using this normal form of the displacement
operator and equations (\ref{k+n})-(\ref{k0n}), the Perelomov coherent states are found to
be \cite{PERL}
\begin{equation}
|\zeta\rangle=(1-|\xi|^2)^k\sum_{s=0}^\infty\sqrt{\frac{\Gamma(n+2k)}{s!\Gamma(2k)}}\xi^s|k,s\rangle.\label{PCN}
\end{equation}

\end{document}